# Effect of shear strain on band structure and electronic properties of phosphorene

Yashasvi Singh Ranawat[1], Rishabh Jain[1]

[1]Department of Ceramic Engineering, Indian Institute of Technology (Banaras Hindu University)

We present an ab-initio investigation of effects of shear strain on band structure and electronic properties of 2D phosphorene. We carried out DFT calculations to determine the shear stress as a function of shear strain and found the monolayer phosphorene has ultimate strength at shear strain 30% and 35% in armchair and zigzag directions, respectively, and it was also found that the monolayer extends in z direction on applying shear strain in both directions. Additionally we derived band structures of phosphorene along both directions under shear strain and have shown that band gap in phosphorene decreases along both directions and that phosphorene shows a semi-metal nature on applying shear strain of magnitude 30% in both directions. The electrical conductivity of phosphorene was estimated by effective mass along zigzag and armchair directions and it is shown that the electrical conductivity is far higher along armchair direction, and that with increasing shear strain conductivity increases along armchair, up to ultimate strength, and zigzag directions.

## I. Introduction

With the rapid miniaturization and cost reduction of electronic appliances there is a tremendous increase in research of 2D materials for their promising applications; many materials have been worked upon: Graphene[1-2], molybdenum disulfide[3], boron nitride[4], etc. One such material, recently researched, is phosphorene. It is obtained by mechanical exfoliation of the black phosphorous, which leads to the formation of monolayer black phosphorous, that has a puckered honeycomb 2D structure, having phosphorous as the only atom[5-6]. It is flexible and also very stable having an inherent and appreciable direct band gap[7-8].

Recent theoretical works have found that on application of axial strain in phosphorene, the band gap has direct-indirect-direct transitions with decrease in the band gap as we increase the axial strain[9-11]. This band gap relation with the axial strain when applied in different direction also shows anisotropic nature of phosphorene. Hence, mechanically induced strain has proven to be an effective mechanism for enhancing the intrinsic properties of phosphorene. As phosphorene posses a 3D structure, with atoms stacked in a puckered sheet with two layers, we attempted to ascertain the effects of shear strain engineering on the band structure of mono-layer phosphorene, hitherto not done, in both zigzag and armchair direction, and determine the effect of shear strain on the electronic properties of phosphorene with the help of effective electron mass calculations.

## II. Simulation Methods

The first-principle ground state calculations were performed using the Quantum-Espresso package[12]. For these DFT calculations[13] the exchange-correlation functional was approximated with general gradient approach (GGA) in Perdew-Burke-Ernzerhof (PBE)[14] type. An energy cut off for wavefunction of 120 Ryd was used to expand wavefunctions into plane waves with 14x10x7 Monkhorst-Pack k grid. Atomic structure was optimised using BFGS method with a force threshold of 0.026 eV/A, calculated by Hellmann–Feynman forces. The simulated relaxed lattice constants were 4.6317 x 3.2983 Å (Illustration 1), and in z axis the cell dimension was taken as 15.919 Å to minimise the interaction between two neighbouring layers to render a 2D system. To simulate the effect of shear strain on the layer of phosphorene, the upper stacked atoms were shifted accordingly to represent the desired shear strain. The structure was then allowed to relax along the z-direction. Band structure calculations were performed using Kohn-Sham electronic band structures.

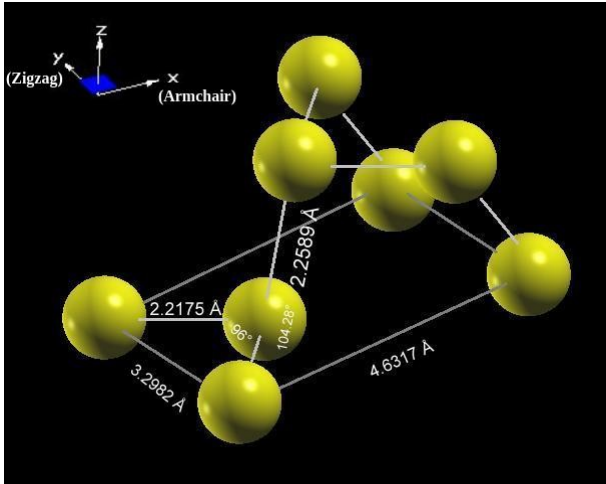

Illustration 1: Structure of Unit cell of relaxed Phosphorene

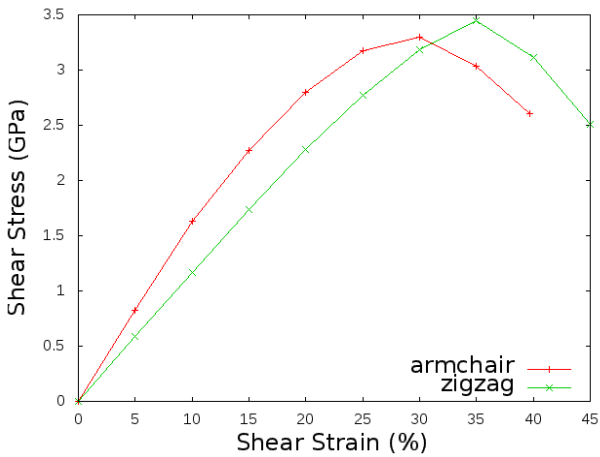

Illustration 2: Graph of Shear stress v/s Shear strain of phosphorene when shear strain is applied along the planar directions.

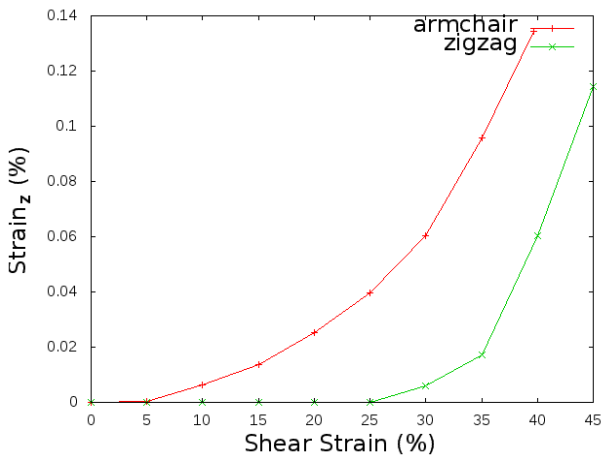

Illustration 3: Graph of Shear strain v/s Shear strain along perpendicular (z) direction of phosphorene when shear strain is applied along the planar directions.

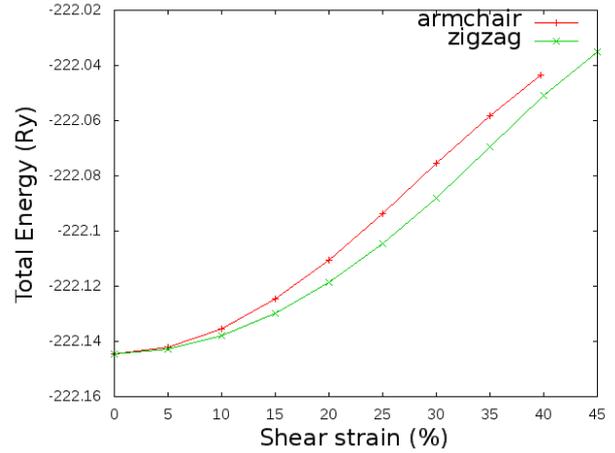

Illustration 4: Graph of total energies of all the structures when shear strain is applied along the planar directions.

We applied these methods to study the topology of the bands, while we were aware that these do not represent true quasiparticle excitations[15].

### III. Results

The shear stress v/s shear strain graph of phosphorene, along armchair and zigzag directions, was plotted (Illustration 2). This curve showed a steady rise in shear stress with strain upto 30% strain in armchair and 35% strain in zigzag direction, where it reaches the maximum. These points represent the shear limit of the material. This showed that phosphorene can sustain shear stress of 3.3 GPa in armchair and 3.45 GPa in zigzag direction. The final strain in z-direction, induced by shear strain, was recorded (Illustration 3). There was an increase in the layer height along z-direction, which was exponential till the shear limit, after which it became linear. Further, the total energy was found to be lowest for the relaxed state (Illustration 4).

The application of shear stress on phosphorene had great impact on its band structure (Illustration 5). The DFT calculation of relaxed structure predicted a band gap of 0.92 eV, which is in agreement with other theoretical work[16,17]. We plotted the band structures along the X-Γ-Y directions with the application of shear strain along armchair and zigzag directions. In the relaxed state, the valence band maximum (VBM) and conduction band minimum (CBM) lied at Γ. When sheared along the armchair direction, the CBM stayed at Γ until the shear strain reached the breaking limit and its value dropped constantly; while the VBM started to shift towards Y direction and its value increased until the shear limit. But when sheared along zigzag direction, the CBM was at Γ until shear strain reached 15%, then shifted towards Y till the shear limit and its value dropped as well; while the VBM stayed at Γ until the shear limit and its value increased.

transition increase was smooth, but abrupt in the zigzag direction at 20% shear strain.

To understand the charge transfer better, the effective mass of electron and hole were calculated along both the directions, with the application of shear strain in those directions. These calculations were done according to formula:

$$m^* = \hbar^2 * \left(\frac{d^2 E}{dk^2}\right)^{-1}$$

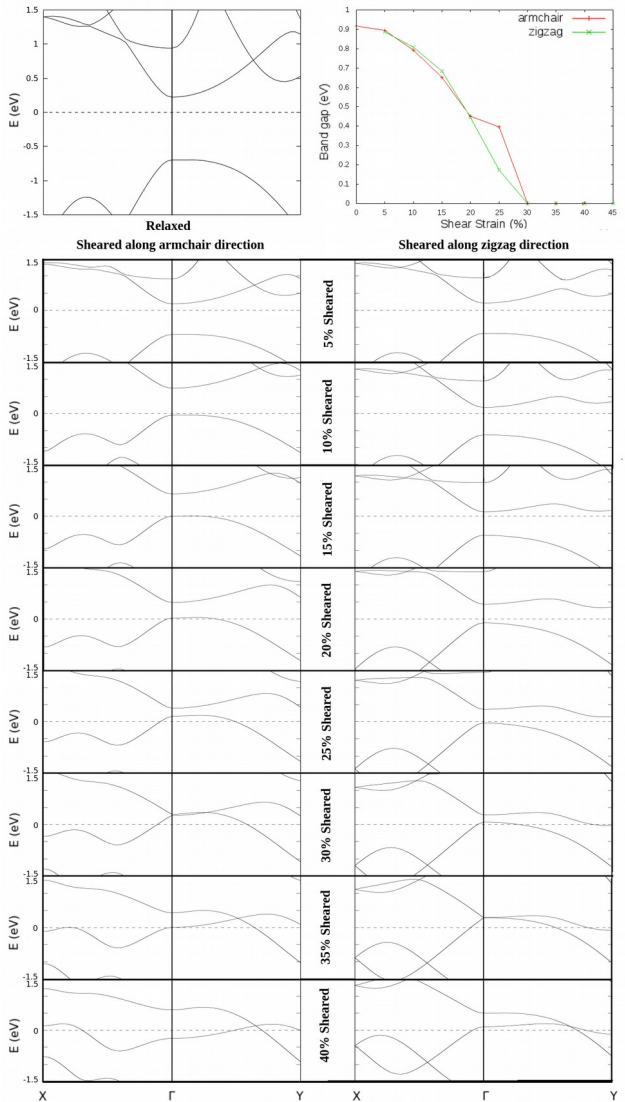

Illustration 5: Band structures of phosphorene when shear strain is applied along the planar directions along with graph of Shear strain applied v/s the band gap.

The band gap of phosphorene decreased with the increase in shear strain in both, armchair and zigzag, directions. In both the directions, the band gap went from being direct to indirect. In the armchair direction, the phonon assisted

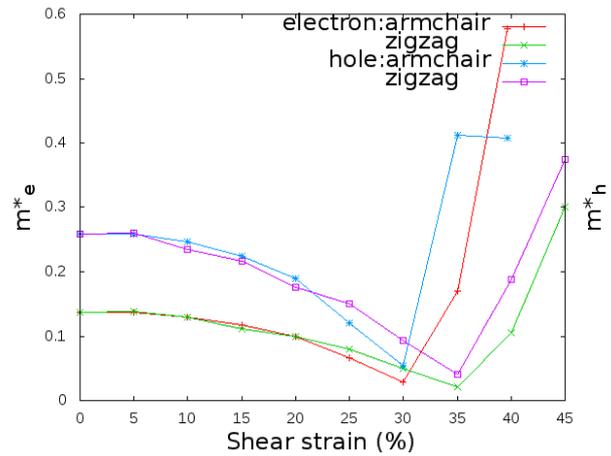

Illustration 6: Graph of Shear strain v/s effective mass of electron and hole when shear strain is applied along the armchair direction.

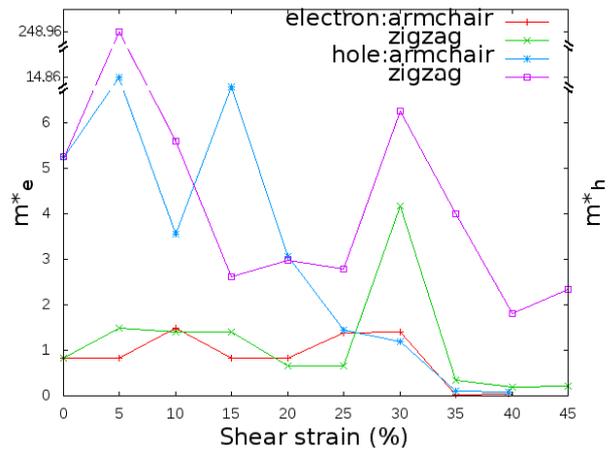

Illustration 7: Graph of Shear strain v/s effective mass of electron and hole when shear strain is applied along the zigza direction.

It was found that the effective mass of electron and hole along the armchair direction was lower that in the zigzag direction. Further, in the armchair direction (Illustration 6), the effective mass of electron and hole dropped till the shear limit was reached when sheared in both the planar directions. In the zigzag direction (Illustration 7), the effective mass of electron and hole were erratic, but did show a general drop. But this value didn't drop down below 1 until shear limit in both directions was reached.

## IV. Conclusion

Using the ab-initio calculations, the effects of shear strain on material structure, band structure, and effective electron and hole masses, of phosphorene, were provided. We found that phosphorene is highly flexible to shear strain as it can withstand shear stress of 3.3 GPa with 30% strain in armchair, and 3.45 GPa with 35% strain in zigzag directions. The band structure was found to be highly tunable; there was a direct-indirect band gap transition, with decreasing energy gap, as shear strain was increased along any of the two directions. Further, we have confirmed the anisotropic nature of phosphorene, in its mechanical and electron properties[6-8]. The effective electron and hole mass calculation indicate high carrier mobility along armchair direction than zigzag direction; which further increases with increasing shear strain. Thus, it can be concluded that phosphorene is a suitable candidate for semi-conductor electronics application, where its properties can be modulated with shear strain application.

**Refrences**